\pdfoutput=1
\documentclass[a4paper,fleqn,usenatbib]{mnras}
\usepackage{graphicx}	% Including figure files
\usepackage{amsmath}	% Advanced maths commands
\usepackage{amssymb}	% Extra maths symbols
\usepackage{bm}
\usepackage{txfonts}

\usepackage[T1]{fontenc}
\usepackage{ae,aecompl}
\usepackage{hyperref}
\usepackage{color}
\title[Gravitational Lens System as a Long Baseline Detector of Extremely Low Frequency PGW]{Gravitational Lens System as a Long Baseline Detector of Extremely Low Frequency Primordial Gravitational Wave}

\author[Wenshuai Liu]{
Wenshuai Liu\thanks{E-mail: 674602871@qq.com}\\
School of Physics, Henan Normal University, Xinxiang 453007, China\\
}

\date{Accepted XXX. Received YYY; in original form ZZZ}

\begin{document}
\label{firstpage}
\pagerange{\pageref{firstpage}--\pageref{lastpage}}
\maketitle

% Abstract of the paper
\begin{abstract}
The effect of extremely low frequency primordial gravitational wave with arbitrary direction of propagation on a gravitational lens system in expanding universe is investigated. From the point of view of real astrophysical lens model, singular isothermal sphere lens model is adopted in the gravitational lens system. The results show that, under the perturbation from extremely low frequency primordial gravitational wave, time delay in the gravitational lens system is very sensitive to extremely low frequency primordial gravitational wave and could strongly deviate from that deduced from theoretical model. This means that the strongly deviate time delay could be the imprint of extremely low frequency primordial gravitational wave on gravitational lens system, indicating that gravitational lens system could be used as a long baseline detector to detect extremely low frequency primordial gravitational wave.
\end{abstract}

% Select between one and six entries from the list of approved keywords.
% Don't make up new ones.

%%%%%%%%%%%%%%%%%%%%%%%%%%%%%%%%%%%%%%%%%%%%%%%%%%

%%%%%%%%%%%%%%%%% BODY OF PAPER %%%%%%%%%%%%%%%%%%

\section{Introduction} \label{sec:intro}
Primordial gravitational waves (PGWs) with a nearly scale-invariant spectrum \citep{1,2,3,4,5,6,7} are a strong prediction of inflation. In addition to leading to a flat, homogeneous, and isotropic Universe, inflation also generates seed perturbations that grow and give rise to large-scale structure in the Universe \citep{8,9,10,11}. Detecting PGWs is crucial for confirming inflation and determining its energy scale. The traditional method of detecting PGWs with extremely low frequency in the range of $10^{-18}$Hz - $10^{-16}$Hz is through the B-modes of polarization of the cosmic microwave background (CMB) \citep{12,13}. However, detecting PGWs with extremely low frequency using B-mode polarization faces challenges due to foreground contamination from dust in our Milky Way. Therefore, it is important to explore alternative observational features induced by such extremely low frequency PGWs.

The research presented in this work suggests that gravitational lens systems could serve as effective detectors of extremely low frequency PGWs. The concept of utilizing gravitational lens systems to detect PGWs with extremely low frequency is proposed based on works \citep{45,46,47}. \cite{45,46} proposed that a gravitational lens system with a point mass lens could be used as a detector of PGWs with extremely low frequency by measuring time delays between different images of a quasar. However, a subsequent study in \cite{47} demonstrated that the approach proposed by \cite{45,46} may not be feasible, as the time delays induced by PGWs cannot be distinguished from the intrinsic time delays caused by the geometry of the gravitational lens system. \cite{48,50} demonstrate that a gravitational lens system with a point mass lens could detect extremely low frequency PGWs when the entire Einstein ring is considered. Both \cite{48,50} and the research in \citep{45,46,47} assume an aligned gravitational lens system with a point mass lens. However, the existence of a point mass lens in strong gravitational lensing is unlikely, and the chances of such an aligned source-deflector-observer configuration occurring in the Universe are very low.

Recently, \cite{43} introduced a new method for detecting extremely low frequency PGWs using a non-aligned gravitational lens system with a point mass lens. The study suggests that the time delay from the gravitational lens system, affected by extremely low frequency PGWs, could deviate from the time delay predicted by the theoretical model by as much as 100 percent. This indicates that gravitational lens systems could potentially serve as long baseline detectors of PGWs with extremely low frequency. However, it's important to note that the direction of propagation of PGWs considered in \cite{43} is not arbitrary, and the lens is modeled as a point mass.

In the general case, the lens is composed of a planar distribution of mass elements. In this study, we focus on the singular isothermal sphere (SIS) lens model, which is commonly used in astrophysics. We investigate the impact of extremely low frequency PGWs with arbitrary propagation directions on a non-aligned gravitational lens system with a SIS lens in an expanding universe. The findings indicate that time delays, when perturbed by extremely low frequency PGWs, could significantly differ from those predicted by the theoretical model. This suggests that gravitational lens systems could potentially be used as long baseline detectors of PGWs. Throughout this work, we adopt $H_0=70km\,s^{-1}\,Mpc^{-1}$, $\Omega_M=0.3$ and $\Omega_\Lambda=0.7$.

\section{Perturbations of PGW on gravitational lens system} \label{sec:basic_eq}
In the gravitational lens system illustrated in Figure 1, the configuration is such that the projection of the source on the lens axis and the observer are not equidistant from the deflector. Specifically, the source, the deflector, and the observer are located at the points ($x=L\beta,y=0,z=-L$), ($x=0,y=0,z=-AL$), and ($x=0,y=0,z=0$) respectively. Here, $A$ represents the ratio of the comoving distance between the observer and the deflector to that between the observer and the source, with the constraint $0<A<1$. Additionally, the speed of light is assumed to be $c=1$, and the metric of the gravitational wave with a wavelength shorter than the horizon of the universe and with arbitrary propagation direction is described as \citep{64,65}.
\begin{equation}
h_{ij}=\frac{a_0}{a}[(u_i u_j-v_i v_j)h_++(u_i v_j+v_i u_j)h_\times]\times\cos(\omega \eta-\mathbf{k} \cdot \mathbf{x})\label{1}
\end{equation}
where $a$ is the scale factor and we set the present value $a_0=1$. $z$ is in form of comoving distance and conformal time. The conformal time is $\eta=t_e+(z+L)$ which could approach the level of approximation, $t_e$ is the time the photons were emitted at $(x=L\beta,y=0,z=-L)$ so that $\omega t_e$ acts as the initial phase, $\mathbf{k}=\omega(\sin\theta\cos\phi, \sin\theta\sin\phi, \cos\theta)$ is the propagation vector, $\mathbf{u}=(\sin\phi, -\cos\phi, 0)$, $\mathbf{v}=(\cos\theta\cos\phi, \cos\theta\sin\phi, -\sin\theta)$, $\omega=2\pi f$, $f$ is the frequency of gravitational wave at present, $h_+$ and $h_\times$ are the amplitude of the two polarizations of the gravitational wave at present, respectively.

\begin{figure*}
 \begin{center}
   \begin{tabular}{cc}
      \includegraphics[width=0.5\textwidth]{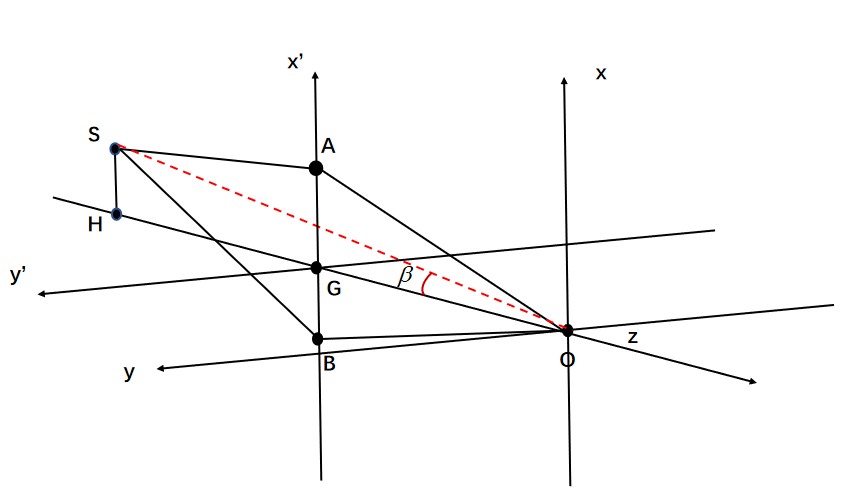}
      \includegraphics[width=0.5\textwidth]{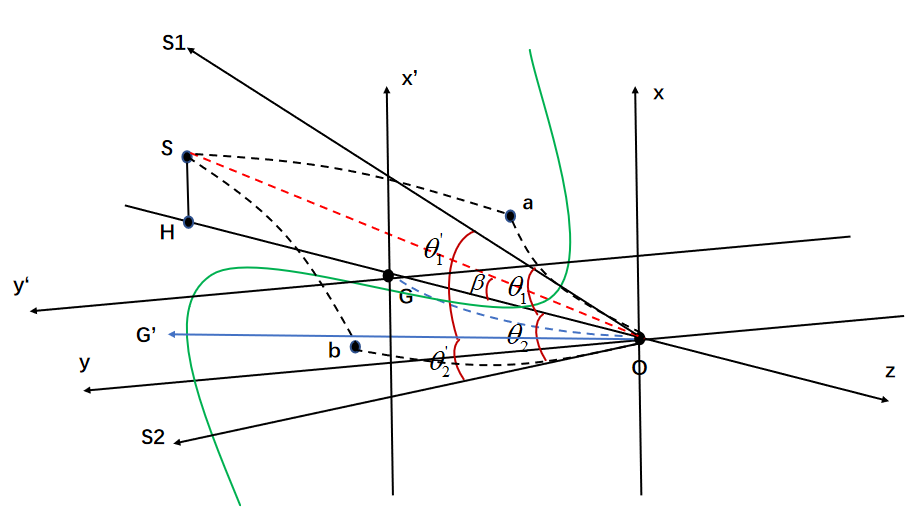}
   \end{tabular}
 \end{center}
	\caption{Left and right show the gravitational lens system. Right is perturbed by extremely low frequency PGW while there is no extremely low frequency PGW in the left. S, G, and O represent the source, the deflector, and the observer, respectively. H is the projection of the source on the line consisted of the observer and the deflector. The green curve in the right represents the extremely low frequency PGW traveling through the gravitational lens system. Under perturbation of extremely low frequency PGW, the dot curves in the right represent the trajectory of light emitted from the source and the deflector. $a$ and $b$ are in the $x'-y'$ plane where $x'$ and $y'$ are parallel to $x$ and $y$, respectively. $\theta_1^{'}$ and $\theta_2^{'}$ represent the angular positions of the two observational image of the source. $\beta$ is the angle between OS and OG.}
    \label{figure1}
\end{figure*}

With a flat Friedmann background perturbed by gravitational potential $U$ of the deflector and extremely low frequency PGW, the total metric in the expanding universe is shown as
\begin{equation}
ds^2=a^2(\eta)[-(1+2U)d\eta^2+(1-2U)(dx^2+dy^2+dz^2)+h_{ij}dx^idx^j]\label{2}
\end{equation}

It shows from \cite{63} that the conformal time delay is equal to the physical time delay to an extremely good approximation. Thus, the following research is in terms of comoving distances and conformal times.

Then we can get the time of travel of light based on the equation given as (the detailed derivation is in Appendix A)
\begin{equation}
T \approx  \int_{-L}^{0}dz[1+\frac{1}{2}\left(\frac{dx}{dz}\right)^2+\frac{1}{2}\left(\frac{dy}{dz}\right)^2
+\frac{1}{2}h_{ij}\frac{dx^i}{dz}\frac{dx^j}{dz}-2U]\label{3}
\end{equation}

In Eq. (\ref{3}), $\frac{dx}{dz}$ and $\frac{dy}{dz}$ should be obtained in order to calculate the time of travel of light perturbed by extremely low frequency PGW with the same method shown in \cite{47}. To get $\frac{dx}{dz}$ and $\frac{dy}{dz}$, it should be noted that we drop the dependence on $x$ and $y$ in $\cos(\omega t-\mathbf{k}\cdot \mathbf{x})$ from Eq. (\ref{1}) with explanation shown in \cite{47} and the fourth term containing $\cos(\omega t-\mathbf{k}\cdot \mathbf{x})$ in Eq. (\ref{3}) is the only place where the dependence on $x$ and $y$ should be included.

Finally, we can obtain $\frac{dx}{dz}$ and $\frac{dy}{dz}$ based on the Euler-Lagrange equation shown as
\begin{equation}
\frac{\partial \mathbb{L}^2}{\partial x^\mu}-\frac{d}{dp}\left(\frac{\partial \mathbb{L}^2}{\dot{x}^\mu}\right)=0\label{4}
\end{equation}
where $\dot{x}^\mu=\frac{dx^\mu}{dp}$, $p$ is the affine parameter for photon and increases monotonically along the worldline, $\mathbb{L}^2$ is expressed as
\begin{eqnarray}
\mathbb{L}^2=a^2(\eta)[-\dot{\eta}^2+\dot{x}^2+\dot{y}^2+\dot{z}^2+h_{11}\dot{x}^2+h_{22}\dot{y}^2+h_{33}\dot{z}^2
\nonumber \\
+2h_{12}\dot{x}\dot{y}+2h_{13}\dot{x}\dot{z}+2h_{23}\dot{y}\dot{z}]
\end{eqnarray}
where the overdot represents $\frac{d}{dp}$ and $\mathbb{L}^2=0$ for photon.

Then, $\frac{dx}{dz}$ and $\frac{dy}{dz}$ are given as follows when $z<-AL$ (the detailed derivation is in Appendix B)
\begin{equation}
\frac{dx}{dz}=\frac{d_2 a_1-d_1a_2}{a_2 b_1-a_1 b_2}+\frac{c_1 a_2-c_2a_1}{a_2 b_1-a_1 b_2}\label{5}
\end{equation}
\begin{equation}
\frac{dy}{dz}=\frac{d_2 b_1-d_1b_2}{a_1 b_2-a_2 b_1}+\frac{c_1 b_2-c_2b_1}{a_1 b_2-a_2 b_1}\label{6}
\end{equation}
where $c_1$ and $c_2$ are integration constants.

In order to obtain $c_1$ and $c_2$, the details are as follows with fixed $\theta$, $\phi$, $A$ and $t_e$. Based on the method of Fermat's principle shown in \cite{47}, after integrating Eq. (\ref{5}) from $z=-L$ to $z=-AL$, the change in the direction of x-axis is
\begin{equation}
\Delta x_1=\int_{-L}^{-AL}\frac{dx}{dz}dz
=e_1+c_1 f_1+c_2 g_1\label{7}
\end{equation}
where $e_1$, $f_1$ and $g_1$ are shown in Appendix C.

Thus,
\begin{equation}
x|_{z=-AL}=L\beta+\Delta x_1\label{8}
\end{equation}
and the change in the direction of y-axis is
\begin{equation}
\Delta y_1=\int_{-L}^{-AL}\frac{dy}{dz}dz=e_2+c_1 f_2+c_2 g_2\label{9}
\end{equation}
where $e_2$, $f_2$ and $g_2$ are shown in Appendix C.

Therefore,
\begin{equation}
y|_{z=-AL}=\Delta y_1\label{10}
\end{equation}

When $z>-AL$, the geodesic equation of light is deflected at $z=-AL$ with a deflected angle $\alpha_x$ in the direction of x-axis and a deflected angle $\alpha_y$ in the direction of y-axis and
\begin{equation}
\alpha_x^2+\alpha_y^2=(2\xi)^2\label{11}
\end{equation}
with relationship
\begin{equation}
\frac{\alpha_x}{\alpha_y}=\frac{x|_{z=-AL}}{y|_{z=-AL}}\label{12}
\end{equation}
which results from the fact that $\mathbf{R}=(x|_{z=-AL},y|_{z=-AL},0)$, $\mathbf{a}=(\frac{dx}{dz}|_{z\rightarrow (-AL)^-},\frac{dy}{dz}|_{z\rightarrow (-AL)^-},1)$ and $\mathbf{b}=(\frac{dx}{dz}|_{z\rightarrow (-AL)^+},$ $\frac{dy}{dz}|_{z\rightarrow (-AL)^+},1)$ are coplanar with relation $(\mathbf{R}\times\mathbf{a})\cdot\mathbf{b}=0$.

In the process above, we are utilizing the singular isothermal sphere (SIS) lens model to represent the lens galaxy. {\color{blue}The density profile of the SIS model is given by $\rho(r)=\frac{\sigma^2}{2\pi Gr^2}$, which is derived by assuming that the matter within the lens behaves as an ideal gas in thermal and hydrostatic equilibrium, confined by a spherically symmetric gravitational potential. Here, $\sigma$ represents the velocity dispersion of the gas particles in the ideal gas or the stellar velocity dispersion of the lens galaxy, and $r$ denotes the distance from the center of the sphere. The deflection angle of light resulting from this lens model is given by $\alpha=4\pi\sigma^2$. Additionally, we set $2\xi=4\pi\sigma^2$.}

The change in the direction of x-axis and y-axis are as follows when $z>-AL$
\begin{equation}
\Delta x_2=\int_{-AL}^0\frac{dx}{dz}dz=h_1+c_1 i_1+c_2 j_1-\alpha_x AL \label{13}
\end{equation}
\begin{equation}
\Delta y_2=\int_{-AL}^0\frac{dy}{dz}dz=h_2+c_1 i_2+c_2 j_2-\alpha_y AL\label{14}
\end{equation}
where $h_1$, $i_1$, $j_1$, $h_2$, $i_2$ and $j_2$ are shown in Appendix C and
\begin{equation}
\frac{dx}{dz}=\frac{d_2 a_1-d_1a_2}{a_2 b_1-a_1 b_2}+\frac{c_1 a_2-c_2a_1}{a_2 b_1-a_1 b_2}-\alpha_x \label{20}
\end{equation}
\begin{equation}
\frac{dy}{dz}=\frac{d_2 b_1-d_1b_2}{a_1 b_2-a_2 b_1}+\frac{c_1 b_2-c_2b_1}{a_1 b_2-a_2 b_1}-\alpha_y \label{21}
\end{equation}
then we get
\begin{equation}
x|_{z=-AL}=-\Delta x_2 \label{15}
\end{equation}
\begin{equation}
\Delta y_1=-\Delta y_2 \label{16}
\end{equation}
\begin{equation}
x|_{z=-AL}^2+\Delta y_1^2=R^2\label{17}
\end{equation}

Combined with Eq. (\ref{11}), Eq. (\ref{12}), Eq. (\ref{15}), and Eq. (\ref{16}), the four unknowns $c_1$, $c_2$, $\alpha_x$, and $\alpha_y$ are solved with two sets of solutions. Inserting $c_1$ and $c_2$ into Eq. (\ref{20}) and Eq. (\ref{21}), we get the angular positions of the two images as
\begin{equation}
\theta_{1x,2x}=-\frac{dx}{dz}|_{z=0}
\end{equation}
\begin{equation}
\theta_{1y,2y}=-\frac{dy}{dz}|_{z=0}
\end{equation}

However, from the point of view of real observation, the angular position of the image of the source is that with respect to the image of the deflector. To get the image angular position of the deflector, we use the same method described above. For the deflector, after integrating Eq. (\ref{5}) and Eq. (\ref{6}) from $z=-AL$ to $z=0$, the change in the direction of x-axis and in the direction of y-axis are
\begin{equation}
\Delta x^{'}=\int_{-AL}^0\frac{dx}{dz}dz=h_1+c_1 i_1+c_2 j_1=0 \label{18}
\end{equation}
\begin{equation}
\Delta y^{'}=\int_{-AL}^0\frac{dy}{dz}dz=h_2+c_1 i_2+c_2 j_2=0 \label{19}
\end{equation}
then the two unknowns $c_1$ and $c_2$ with one set of solution are obtained. Finally, the angular position of the image of the deflector is given as follows after inserting $c_1$ and $c_2$ into Eq. (\ref{5}) and Eq. (\ref{6})
\begin{equation}
\theta_{dx}=-\frac{dx}{dz}|_{z=0}
\end{equation}
\begin{equation}
\theta_{dy}=-\frac{dy}{dz}|_{z=0}
\end{equation}

Thus, from the point of view of real observation, the image angular position of the source, which is the image of the source with respect to the image of the deflector, is shown as
\begin{equation}
\theta^{'}_{1x,2x}=\theta_{1x,2x}-\theta_{dx} \label{22}
\end{equation}
\begin{equation}
\theta^{'}_{1y,2y}=\theta_{1y,2y}-\theta_{dy} \label{23}
\end{equation}

Finally, the time delay $\Delta T_{Observation}$ based on Eq. (\ref{3}) can be calculated and the theoretical time delay $\Delta T_{Theory}$ is obtained as following with the image positions shown in Eq. (\ref{22}) and Eq. (\ref{23})
\begin{equation}
\Delta T_{Theory}=2\eta AL(-\theta_1^{'}+\theta_2^{'}) \label{27}
\end{equation}
and $-\int_{-L}^0 2U_1dz+\int_{-L}^0 2U_2dz$ in $\Delta T_{Observation}$ could be expressed as
\begin{eqnarray}
-\int_{-L}^0 2U_1dz+\int_{-L}^0 2U_2dz&=&2\eta AL(-\frac{R_1}{AL}+\frac{R_2}{AL})
\nonumber \\
&=&2\eta(-R_1+R_2) \label{28}
\end{eqnarray}
then, the obtained $\kappa=\frac{\Delta \rm T_{\rm {Theory}}-\Delta \rm T_{\rm {Observation}}}{\Delta \rm T_{\rm {Theory}}}$ could be the hint of extremely low frequency PGW.

\begin{figure*}
 \begin{center}
   \begin{tabular}{cc}
      \includegraphics[width=0.33\textwidth]{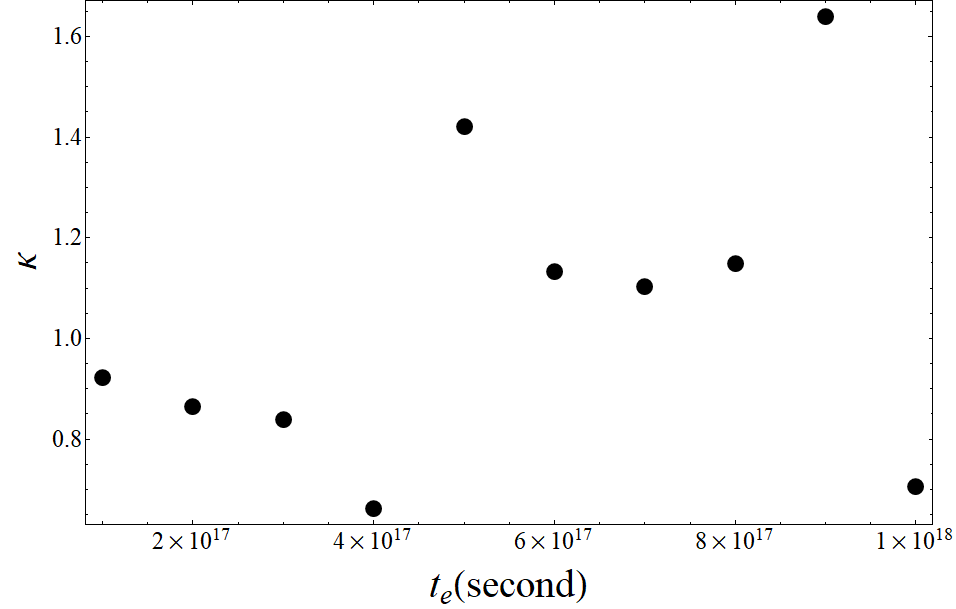}
      \includegraphics[width=0.33\textwidth]{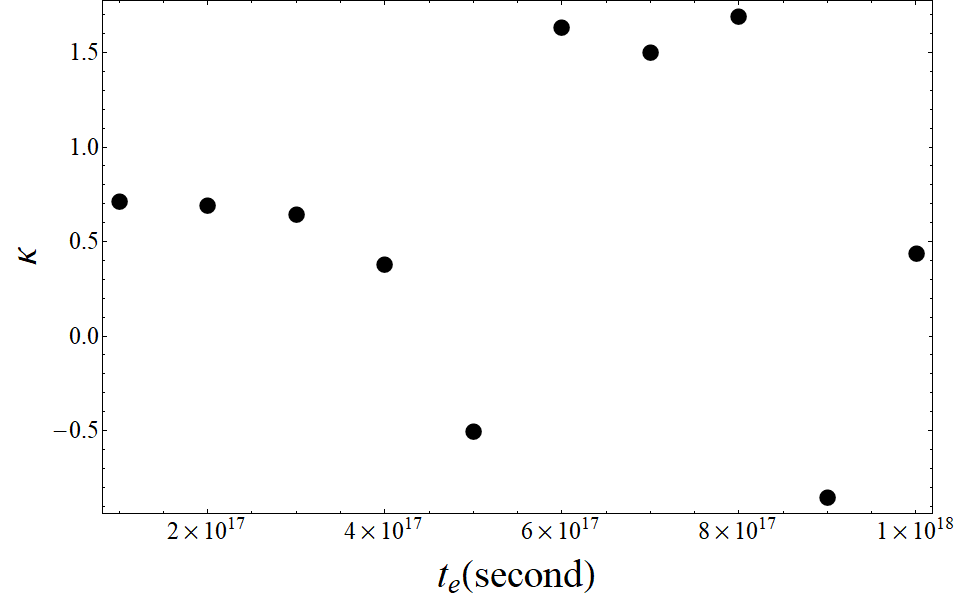}
      \includegraphics[width=0.33\textwidth]{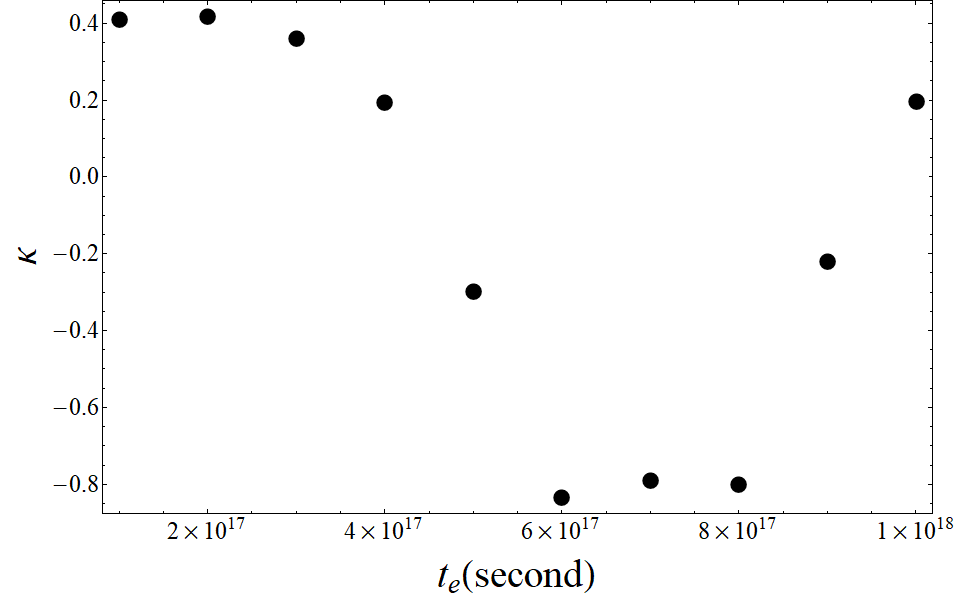}\\
      \includegraphics[width=0.33\textwidth]{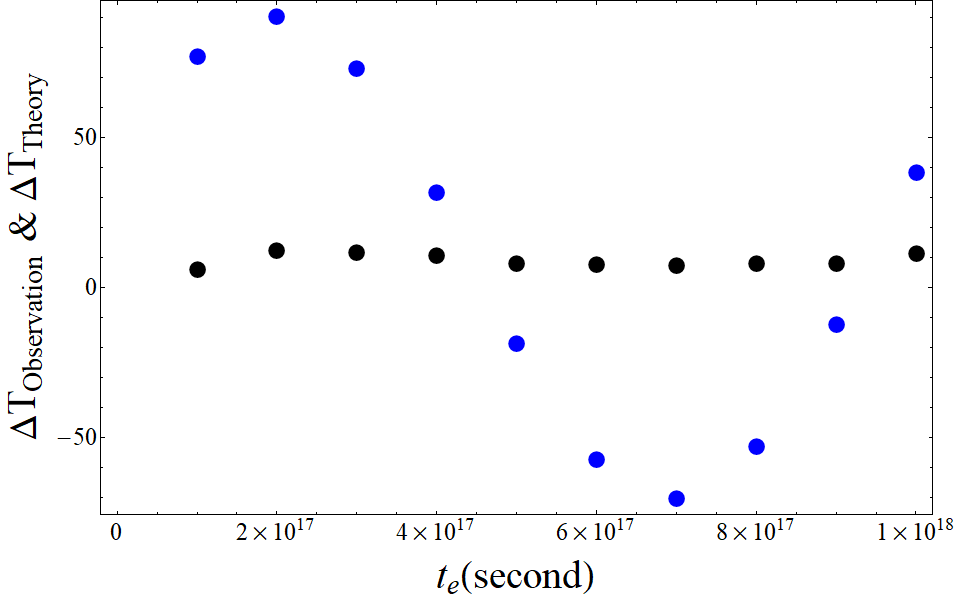}
      \includegraphics[width=0.33\textwidth]{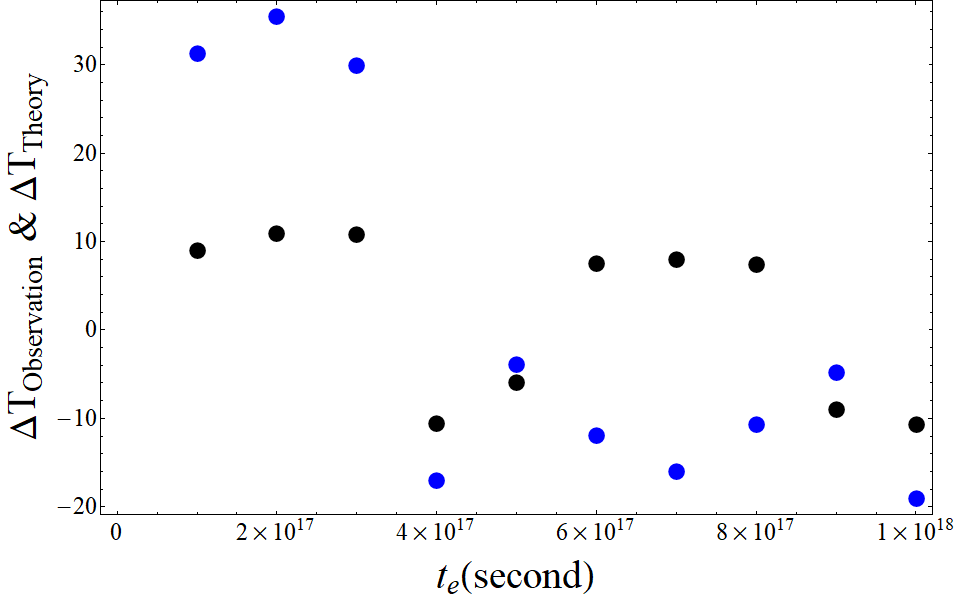}
      \includegraphics[width=0.33\textwidth]{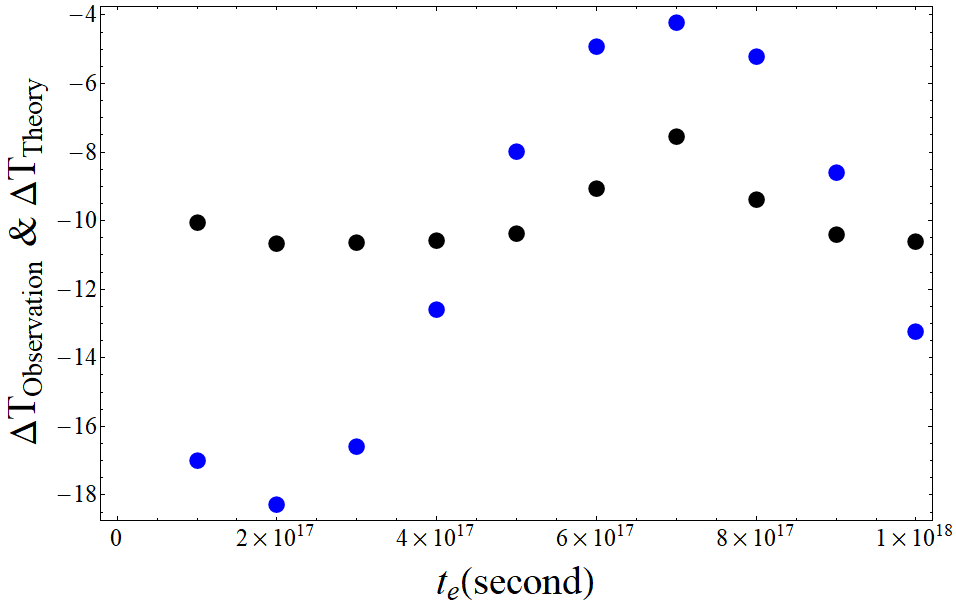}
   \end{tabular}
 \end{center}
	\caption{Up and bottom are $\kappa$ and time delay in unit of day along with $t_e$. $t_e$ is the time the photons were emitted from the source. Left, middle and right are the results when $h=1.45\times10^{-7}, 4.58\times10^{-8}, 1.45\times10^{-8}$, respectively. The black dots and the blue dots on the bottom represent the observational time delay $\Delta T_{Observation}$ and the theoretical time delay $\Delta T_{Theory}$, respectively. The ratio of the comoving distance between the observer and the deflector to that between the observer and the source is $A=0.2$.}
    \label{figure6}
\end{figure*}

\begin{figure*}
 \begin{center}
   \begin{tabular}{cc}
      \includegraphics[width=0.33\textwidth]{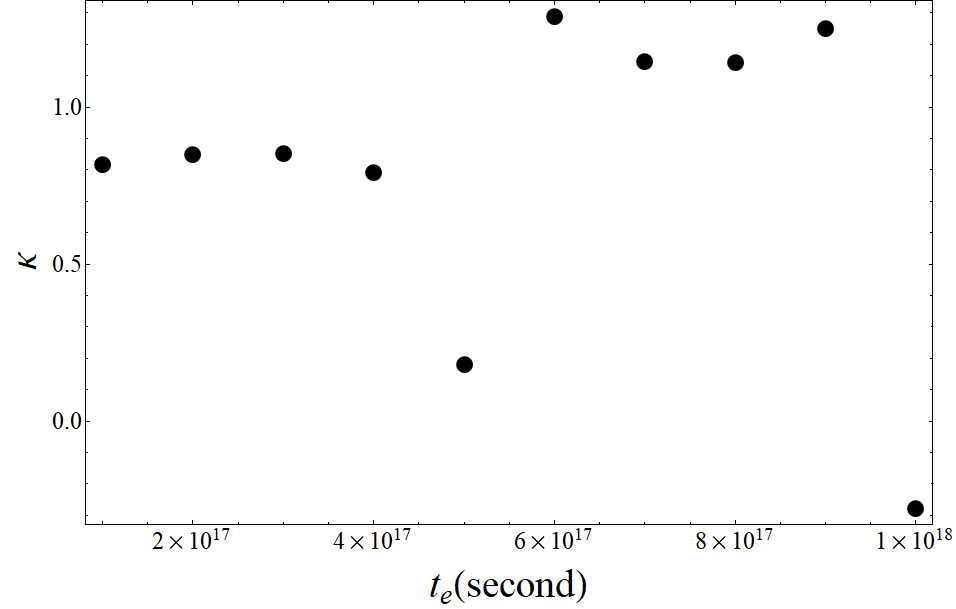}
      \includegraphics[width=0.33\textwidth]{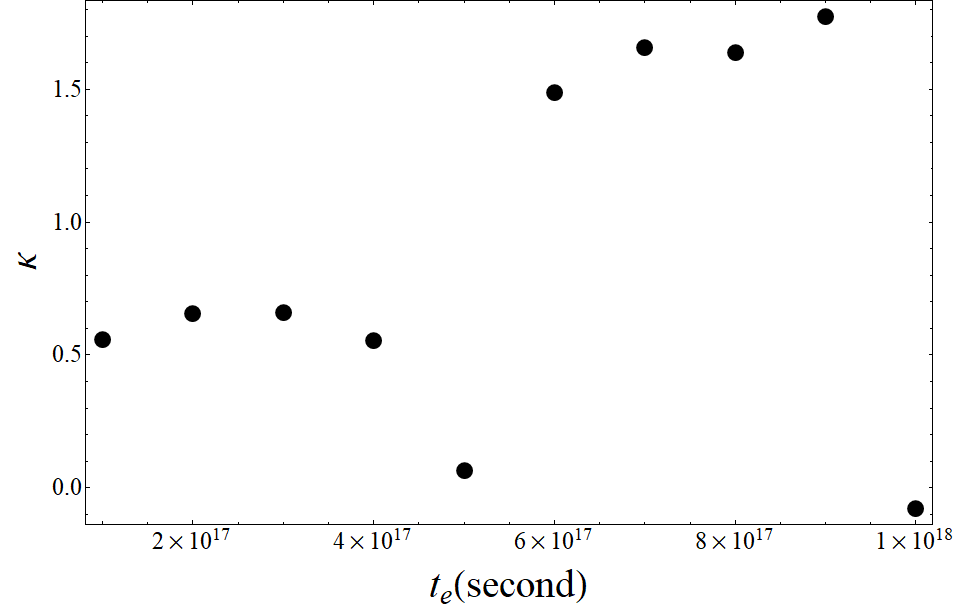}
      \includegraphics[width=0.33\textwidth]{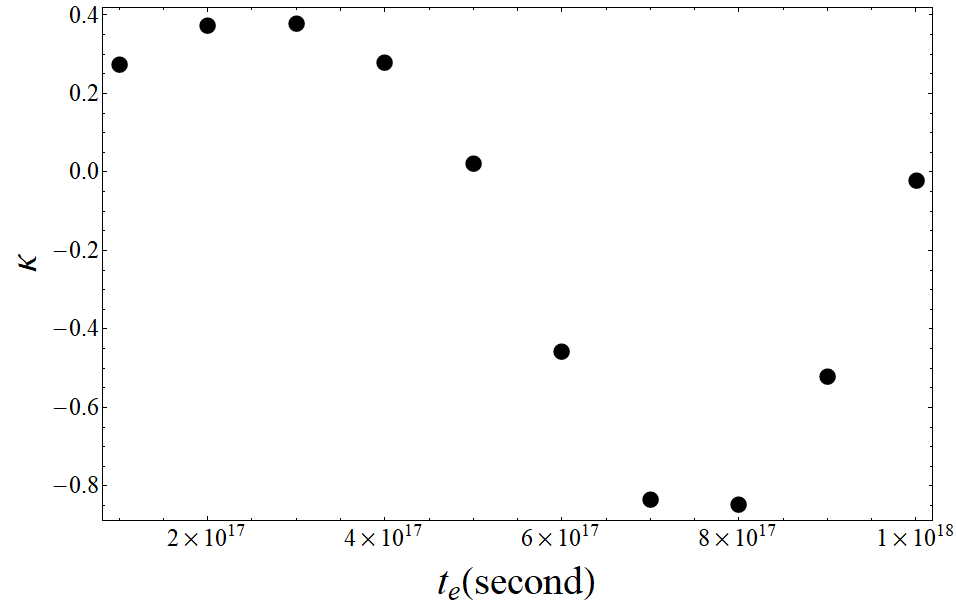}\\
      \includegraphics[width=0.33\textwidth]{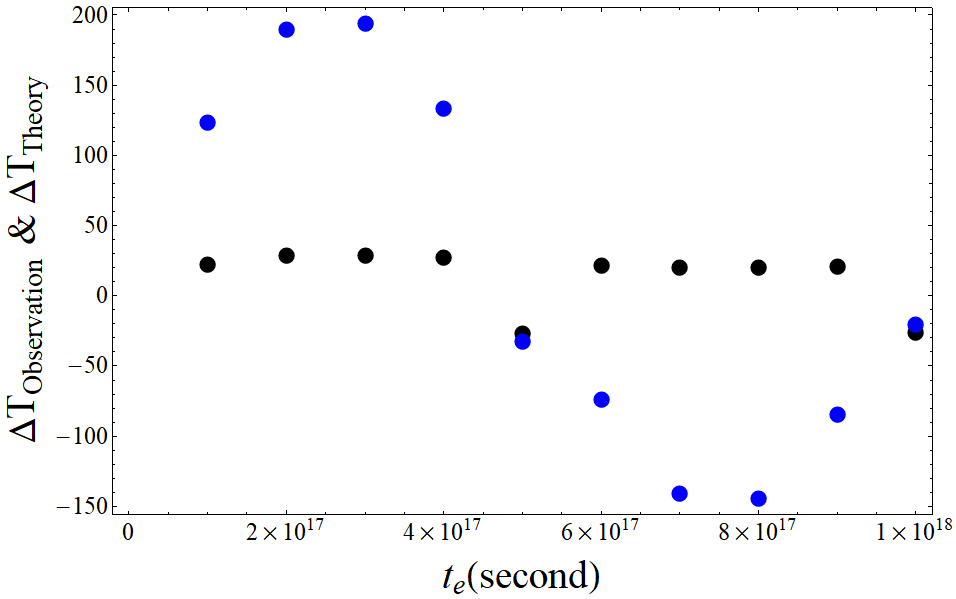}
      \includegraphics[width=0.33\textwidth]{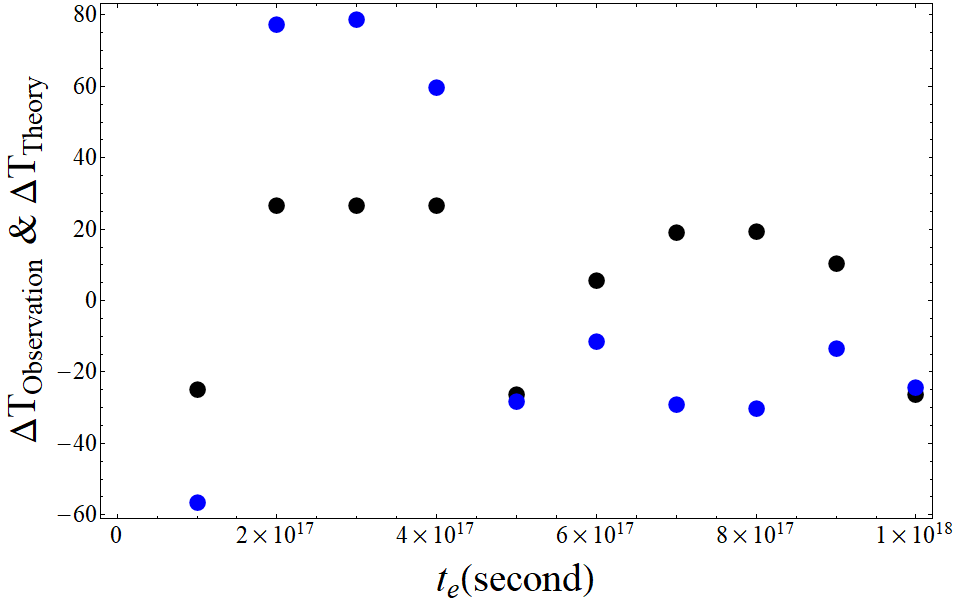}
      \includegraphics[width=0.33\textwidth]{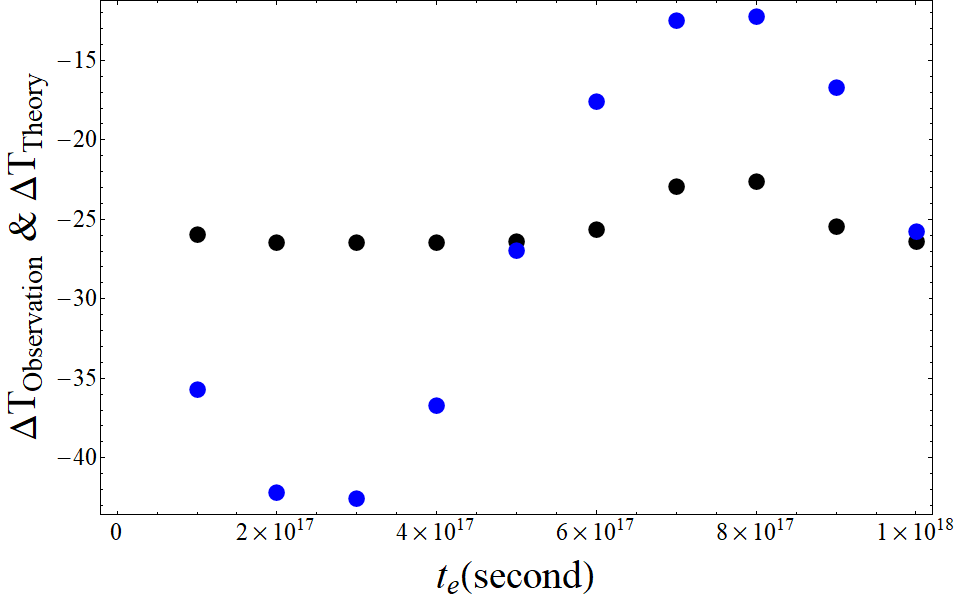}
   \end{tabular}
 \end{center}
	\caption{Up and bottom are $\kappa$ and time delay in unit of day along with $t_e$. $t_e$ is the time the photons were emitted from the source. Left, middle and right are the results when $h=1.45\times10^{-7}, 4.58\times10^{-8}, 1.45\times10^{-8}$, respectively. The black dots and the blue dots on the bottom represent the observational time delay $\Delta T_{Observation}$ and the theoretical time delay $\Delta T_{Theory}$, respectively. The ratio of the comoving distance between the observer and the deflector to that between the observer and the source is $A=0.5$.}
    \label{figure6}
\end{figure*}

\begin{figure*}
 \begin{center}
   \begin{tabular}{cc}
      \includegraphics[width=0.5\textwidth]{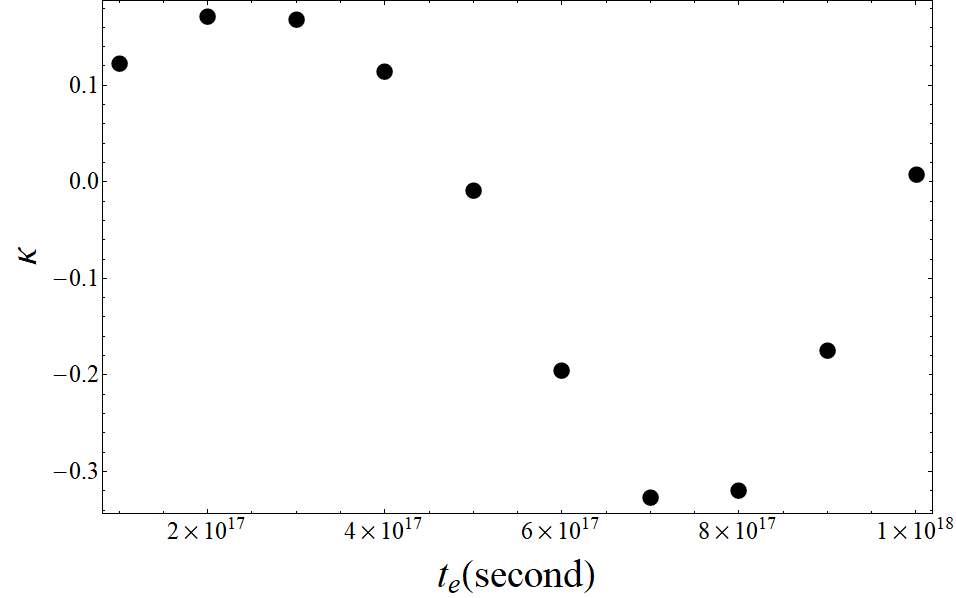}
      \includegraphics[width=0.5\textwidth]{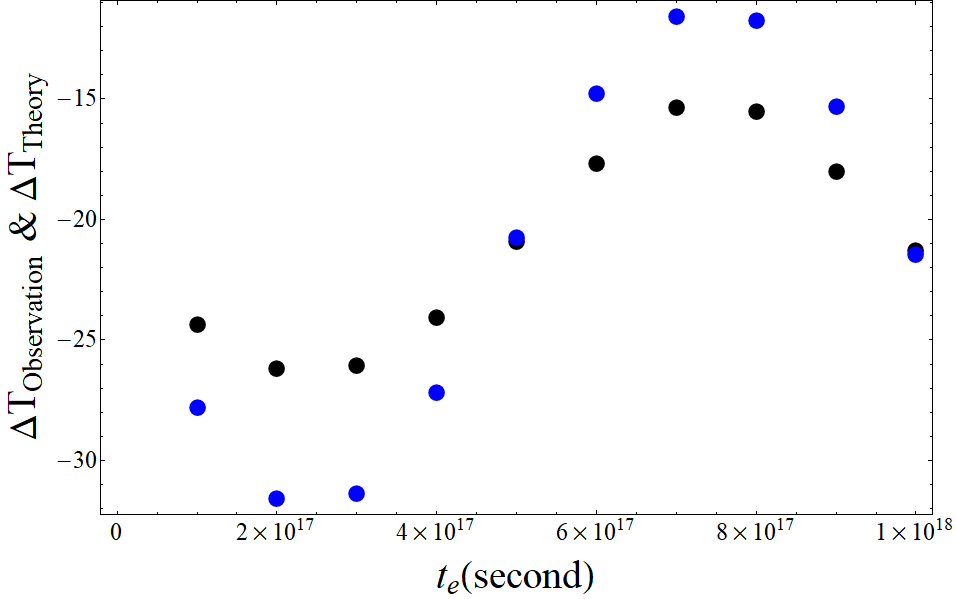}
   \end{tabular}
 \end{center}
	\caption{Left and right are $\kappa$ and time delay in unit of day along with $t_e$. $t_e$ is the time the photons were emitted from the source. The black dots and the blue dots on the right represent the observational time delay $\Delta T_{Observation}$ and the theoretical time delay $\Delta T_{Theory}$, respectively.}
    \label{figure6}
\end{figure*}

The calculations based on Eq. (\ref{22}) and Eq. (\ref{23}) show that the positions of the two source images and the position of the lens image are collinear, thus, we set $\theta'_1=\sqrt{\theta'^{2}_{1x}+\theta'^{2}_{1y}}$ and $\theta'_2=\sqrt{\theta'^{2}_{2x}+\theta'^{2}_{2y}}$ which represent the positions of the two observational images.

In the following calculations, we set $L=8.4$Gpc in form of comoving distance (corresponding to redshift equal to 6.4) and $A=0.2,0.5$. The velocity dispersion in lens galaxy discovered is usually about several hundred $km/s$, for instance, $\sigma\approx 323 km/s$ in the len galaxy in RXJ1131-1231 \citep{51}. We set $\xi=6.79\times 10^{-6}$ corresponding to $\sigma\approx 310 km/s$ in this work and we take $\beta=2\times10^{-7}$. $A$, $\xi$, $\beta$ and $L$ can take other values and are not limited to the ones adopted here.

The primordial spectrum of primordial gravitational wave is defined far outside the horizon during inflation as \citep{59}
\begin{equation}
h(k)=\Delta_t (k)
=\Delta_{R}(k_{0}) \, r^{1/2}(\frac{k}{k_{0}})^{\frac{n_t}{2}
+\frac{1}{4}\alpha_t \ln(\frac{k}{k_0})} \label{39}
\end{equation}
where $k_{0}$ is a pivot conformal wave number and corresponds to a physical wave number
through $k_0/a(\tau_H) = 0.05 Mpc^{-1}$, $\Delta_{R}^{2}(k_0)=2.1\times10^{-9}$ is the curvature perturbation from observation \citep{67}, and $r\equiv \Delta^2_{t}(k_0)/\Delta^2_{R}(k_0)$ is the tensor-scalar ratio and the recent constrain on $r$ is $r< 0.032$ \citep{66}.

It shows that $n_t$ and $\alpha_t$ are near $0$ \citep{62,59,61,60} and the primordial spectrum is nearly flat in the extremely low frequency range. According to the research \citep{62,59,61,60}, the amplitude of the PGW at a frequency of $f=10^{-18}Hz$ at present is about one order of magnitude lower than the primordial value. In the following, we set the tensor-scalar ratio $r$ far lower than the current upper limit to see whether or not PGW at a frequency of $f=10^{-18}Hz$ could affect the gravitational lens system.

To demonstrate the impact of PGW with a frequency of $f=10^{-18}Hz$ on a gravitational lens system with a much lower tensor-scalar ratio than the current upper limit, we set $r=0.001,0.0001,0.00001$. Subsequently, we find that the primordial amplitude of the PGW at a frequency of $f=10^{-18}Hz$ is $h=1.45\times10^{-6}, 4.58\times10^{-7}, 1.45\times10^{-7}$ according to Eq. (\ref{39}), respectively. Consequently, the current amplitude of the PGW at a frequency of $f=10^{-18}Hz$ is $h=1.45\times10^{-7}, 4.58\times10^{-8}, 1.45\times10^{-8}$ when $r=0.001,0.0001,0.00001$, respectively. The corresponding $h_+$ and $h_\times$ values are $h_+=\frac{h}{\sqrt{2}}$ and $h_\times=\frac{h}{\sqrt{2}}$, respectively.

When $\theta \ne \frac{\pi}{2}$ and $\phi \ne 0$, only numerical results are available. For simplicity, let's set $\theta=1.4$ and $\phi=0.2$. Using $h=1.45\times10^{-7}, 4.58\times10^{-8}, 1.45\times10^{-8}$, $f=10^{-18}$ Hz and $A=0.2$, we obtain $\kappa$ as well as observational and theoretical time delays in Figure 2. It shows from Figure 2 that the observed time delay may deviate significantly from the predicted theoretical value. The results when $A=0.5$ are displayed in Figure 3 where the disparity between the observed and theoretical time delays could be as large as 150 days, indicating that PGW with a frequency of $f=10^{-18}$ Hz could have a significant impact on gravitational lens systems. It's important to note that the results hold when the projection of the source is on the $y-z$ plane and not limited to the z-axis.

When the frequency of PGW is $f=10^{-19}$Hz, its wavelength is larger than the horizon of the universe, resulting that such PGW is frozen and does't evolve. Only when the horizon size again is larger than its wavelength can it start oscillate and propagate. When $f=10^{-19}$Hz, $\eta$ in Eq. (\ref{1}) is $\eta=t_e$. Then, with $\theta=1.4$ and $\phi=0.2$, $h=1.45\times10^{-6}$, $f=10^{-19}$Hz and $A=0.2$, $\kappa$ along with $t_e$ and the time delays in unit of day along with $t_e$ are shown in Figure 4. {\color{blue}When $h=4.58\times10^{-7}, 1.45\times10^{-7}$, the calculated $\kappa$ is much closer to 0 and the resulting $\Delta T_{Theory}$ is closer to $\Delta T_{Observation}$ compared with that obtained based on $h=1.45\times10^{-6}$, thus, these are not shown in Figure 4.}

In the process above, the lens model used to calculate the theoretical time delay $\Delta T_{Theory}$ is the SIS model we adopt during the calculation of the observational time delay $\Delta T_{Observation}$ for simplicity and the explanation is as follows. In practice, the lens should be reconstructed based on the source's images in order to calculate the theoretical time delay, which is beyond the scope of this work. In the case of a single source in the gravitational lens system, a transformation exists that leaves the observable images unchanged but produces a family of different mass distributions, known as the mass-sheet degeneracy \citep{80}. To break this degeneracy, at least two sources at different redshifts are needed in the gravitational lens system \citep{81}. In this work, we assume the given SIS lens is in a gravitational lens system with multiple sources of different redshifts. The source used to calculate $\kappa$ is at a high redshift (redshift equal to 6.4), while the other two sources are at low redshifts (less than 0.6 when $A=0.2$, for example). Based on the images of the two sources with low redshift, the lens model could be reconstructed accurately. For the source with low redshift, the perturbation of extremely low frequency PGWs on the images of the low redshift source can be negligible and therefore this is not shown in this work. As a result, the reconstructed lens model could be consistent with the given SIS model. Thus, we use the given SIS lens model as the reconstructed lens model to calculate the theoretical time delay $\Delta T_{Theory}$ between the images of the source with high redshift.

In real observations, the time delay in observations due to dedicated and long-duration monitoring has uncertainties of a few percent \citep{53,54}. Furthermore, deep high-resolution imaging of the lensed arc could help constrain the potential difference between different images at a few percent \citep{55}. To resolve lensing degeneracy caused by the distribution of mass in the lens galaxy, it is crucial to obtain the lens galaxy's stellar velocity dispersion. Additionally, the weak lensing effect along the line of sight due to galaxies close in projection to the gravitational lens system needs to be inferred with independent observational data and could be constrained at about a 5$\%$ level \citep{55}. The uncertainties of all these sources are about 5-8$\%$ precision \citep{51}. Moreover, substructure in a precise model of the deflector potential could produce perturbations of a fraction of one day \citep{56}, and gravitational microlensing leads to perturbations on the order of days if the source in the gravitational lens system is a quasar \citep{57}, with no perturbations expected if the source is transient, such as a binary neutron star merger. When the time delay anomaly, typically observed in gravitational lensing, is larger than that produced by the total perturbations described above, it may indicate the existence of extremely low frequency PGWs.

It shows from \cite{52} that the observational time delay is about a year while the time delay from prediction based on their mass model is about 410 days, thus, microlensing and substructure in the lens can't explain the difference of several tens of days. If other perturbations including the potential difference between different images, lens galaxy mass distribution, and line of sight weak lensing effect are inferred with independent observational data and such inferred perturbations couldn't reach several tens of days, there may exist the imprint of extremely low frequency PGWs on the gravitational lens system.

PGWs are isotropic statistically, meaning that the amplitude of PGW depends only on the frequency and not on the direction of propagation. Thus, we can average $\kappa^2$ over phase $\delta=\omega t_e$ and over $4\pi$ solid angle in the direction of $\mathbf{k}$ to obtain the mean square as
\begin{equation}
\langle\kappa^2\rangle=\int^{2\pi}_{0}\frac{d\delta}{2\pi}\int^{2\pi}_{0}\frac{d\phi}{2\pi}\int^{\pi}_{0}\frac{\sin\theta}{2}d\theta\kappa^2 \label{25}
\end{equation}

We get an approximation of $\sqrt{\langle\kappa^2\rangle}$ by removing $\frac{a_0}{a}$ from Eq. (\ref{1}) in order to save computational time, and by dividing $\phi\in[-0.528,-0.528+2\pi]$, $\theta\in[-0.528,-0.528+\pi]$ and $\omega t_e\in[0,2\pi]$ into ten, five and ten equal intervals, respectively. For $h=1.45\times10^{-7}, 4.58\times10^{-8}, 1.45\times10^{-8}$ when $f=10^{-18}Hz$, the resulting $\sqrt{\langle\kappa^2\rangle}$ are $0.65$, $0.29$ and $0.09$, respectively. It should be noted that these approximations of $\sqrt{\langle\kappa^2\rangle}$ are lower than the true value of $\sqrt{\langle\kappa^2\rangle}$ when the whole Eq. (\ref{1}) is considered due to the fact that $\sqrt{\langle\kappa^2\rangle}$ increases as $h_{ij}$ increases when $a$ decreases as redshift increases.

In order to get the expectation value of $\sqrt{\langle\kappa^2\rangle}$, we get the following with the same method as that in \citep{42,49}
\begin{equation}
\sqrt{\langle\kappa^2\rangle}^{'}=\int^{+\infty}_{0}\frac{\sqrt{\langle\kappa^2\rangle}}{f}df \label{26}
\end{equation}

In general, a large $|\kappa|$ for a SIS lens is the hint of extremely low frequency PGWs.

\section{Conclusions and Discussions} \label{sec:discussion}
This work investigates the impact of extremely low frequency PGWs on the time delay between different images of a source in a gravitational lens system with a singular isothermal sphere (SIS) lens model. The results show that the time delay between images could be significantly affected by extremely low frequency PGWs, leading to noticeable deviations from the theoretical model. This suggests that gravitational lens systems with SIS lens models could potentially be used as long baselines to detect extremely low frequency PGWs. Thus, in addition to B-mode polarization, the observable deviations in time delays in gravitational lensing could serve as an alternative feature induced by extremely low frequency PGWs.

It should be noted that the conclusion we've reached applies not only to a specific type of deflector (SIS gravitational deflector) but also to general lens models. Whether we consider the specific lens model used in this study or a general one, the potential existence of extremely low frequency PGWs could be confirmed if the observed time delay significantly differs from that predicted by theory with perturbations described above, such as a precise model of the deflector potential across the images, the distribution of mass in the lens galaxy, weak lensing effects along the line of sight, substructure within the lens, and gravitational microlensing. In other words, if the total perturbations inferred from independent observational data cannot explain the observed time delay anomaly, extremely low frequency PGWs may exist. This study has utilized a single gravitational lens system to detect the hint of extremely low frequency PGWs. To reduce systematic uncertainty, future research may explore the use of a Gravitational Lensing Array (GLA) comprising numerous gravitational lens systems to detect the possible existence of extremely low frequency PGWs. However, this is not within the scope of the current study and will be investigated in the future.

\section*{DATA AVAILABILITY}
The data underlying this article will be shared on reasonable request to the corresponding author.

%\bibliographystyle{aasjournal}
%\bibliography{reference} % if your bibtex file is called example.bib

\section*{Appendix a}
With the metric of expanding universe and the perturbation from primordial gravitational wave
\begin{equation}
ds^2=-a^2({\eta})(1+2U)d\eta^2+a^2({\eta})(1-2U)(dx^2+dy^2+dz^2)+a^2({\eta})h_{ij}dx^idx^j\label{101}
\end{equation}
for photon, we get
\begin{equation}
d\eta=(1+2U)^{-\frac{1}{2}}\frac{1}{a}(\delta_{ij}a^2(1-2U)dx^idx^j+a^2h_{ij}dx^idx^j)^\frac{1}{2}\label{102}
\end{equation}

Combined with
\begin{equation}
dl=\sqrt{\delta_{ij}dx^idx^j}\label{103}
\end{equation}
Eq. (\ref{102}) changes to
\begin{equation}
d\eta=(1+2U)^{-\frac{1}{2}}((1-2U)+h_{ij}\frac{dx^i}{dl}\frac{dx^j}{dl})^\frac{1}{2}dl\label{104}
\end{equation}

With
\begin{equation}
\frac{dz}{dl}=\cos{\frac{\sqrt{dx^2+dy^2}}{dz}}\approx1-\frac{1}{2}\left(\frac{dx}{dz}\right)^2-\frac{1}{2}\left(\frac{dy}{dz}\right)^2\label{105}
\end{equation}
we have
\begin{equation}
dl=(1+\frac{1}{2}\left(\frac{dx}{dz}\right)^2+\frac{1}{2}\left(\frac{dy}{dz}\right)^2)dz\label{106}
\end{equation}
Then
\begin{equation}
d\eta=dz(1+\frac{1}{2}\left(\frac{dx}{dz}\right)^2+\frac{1}{2}\left(\frac{dy}{dz}\right)^2+\frac{1}{2}h_{ij}\frac{dx^i}{dz}\frac{dx^j}{dz}-2U)\label{108}
\end{equation}
Finally, we get
\begin{equation}
T=\int_{-L}^{0}dz[1+\frac{1}{2}\left(\frac{dx}{dz}\right)^2+\frac{1}{2}\left(\frac{dy}{dz}\right)^2\label{109}
+\frac{1}{2}h_{ij}\frac{dx^i}{dz}\frac{dx^j}{dz}-2U]
\end{equation}
where the relationship between $a$ and $z$ can be obtained by
\begin{equation}
a=\frac{1}{1+Z}
\end{equation}
and
\begin{equation}
-z=\int_{0}^{Z}\frac{dZ'}{H_0\sqrt{\Omega_M(1+Z')^3+\Omega_\Lambda}}
\end{equation}
in which $Z$ is redshift.

\section*{Appendix b}
Based on
\begin{equation}
\frac{\partial \mathbb{L}^2}{\partial x^\mu}-\frac{d}{dp}\left(\frac{\partial \mathbb{L}^2}{\dot{x}^\mu}\right)=0\label{120}
\end{equation}
and
\begin{equation}
\mathbb{L}^2=a^2(-\dot{\eta}^2+\dot{x}^2+\dot{y}^2+\dot{z}^2+h_{11}\dot{x}^2+h_{22}\dot{y}^2+h_{33}\dot{z}^2+2h_{12}\dot{x}\dot{y}+2h_{13}\dot{x}\dot{z}+2h_{23}\dot{y}\dot{z})\label{121}
\end{equation}
we get
\begin{eqnarray}
&&a^2\omega(h_{11}\dot{x}^2+h_{22}\dot{y}^2+h_{33}\dot{z}^2+2h_{12}\dot{x}\dot{y}+2h_{13}\dot{x}\dot{z}+2h_{23}\dot{y}\dot{z})\tan(\omega \eta-\mathbf{k} \cdot \mathbf{x})
\nonumber \\
&&+\frac{da^2}{d\eta}\frac{\mathbb{L}^2}{a^2}=\frac{d[a^2(2\dot{\eta})]}{dp} \label{122} \\
\nonumber \\
&&a^2\sin\theta\cos\phi\omega(h_{11}\dot{x}^2+h_{22}\dot{y}^2+h_{33}\dot{z}^2+2h_{12}\dot{x}\dot{y}+2h_{13}\dot{x}\dot{z}
\nonumber \\
&&+2h_{23}\dot{y}\dot{z})\tan(\omega \eta-\mathbf{k} \cdot \mathbf{x})=\frac{d[2a^2(\dot{x}h_{11}+\dot{y}h_{12}+\dot{z}h_{13}+\dot{x})]}{dp} \label{123} \\
\nonumber \\
&&a^2\sin\theta\sin\phi\omega(h_{11}\dot{x}^2+h_{22}\dot{y}^2+h_{33}\dot{z}^2+2h_{12}\dot{x}\dot{y}+2h_{13}\dot{x}\dot{z}
\nonumber \\
&&+2h_{23}\dot{y}\dot{z})\tan(\omega \eta-\mathbf{k} \cdot \mathbf{x})=\frac{d[2a^2(\dot{y}h_{22}+\dot{x}h_{12}+\dot{z}h_{23}+\dot{y})]}{dp} \label{124} \\
\nonumber \\
&&a^2\cos\theta\omega(h_{11}\dot{x}^2+h_{22}\dot{y}^2+h_{33}\dot{z}^2+2h_{12}\dot{x}\dot{y}+2h_{13}\dot{x}\dot{z}
\nonumber \\
&&+2h_{23}\dot{y}\dot{z})\tan(\omega \eta-\mathbf{k} \cdot \mathbf{x})=\frac{d[2a^2(\dot{z}h_{33}+\dot{x}h_{13}+\dot{y}h_{23}+\dot{z})]}{dp} \label{125}
\end{eqnarray}
Combined with Eq. (\ref{122}) to Eq. (\ref{125}), we have
\begin{equation}
a_1^{'}\dot{y}+b_1^{'}\dot{x}+d_1^{'}\dot{z}=c_1 \label{126}
\end{equation}
\begin{equation}
a_2^{'}\dot{y}+b_2^{'}\dot{x}+d_2^{'}\dot{z}=c_2\label{127}
\end{equation}
where $c_1$ and $c_2$ are constants and
\begin{eqnarray}
a_1^{'}&=&a^2(h_{22}\cos\phi+\cos\phi-h_{12}\sin\phi)\\
a_2^{'}&=&a^2(h_{22}\cos\theta+\cos\theta-h_{23}\sin\theta\sin\phi)\\
b_1^{'}&=&a^2(h_{12}\cos\phi-h_{11}\sin\phi-\sin\phi)\\
b_2^{'}&=&a^2(h_{12}\cos\theta-h_{13}\sin\theta\sin\phi)\\
d_1^{'}&=&a^2(h_{23}\cos\phi-h_{13}\sin\phi)\\
d_2^{'}&=&a^2(h_{23}\cos\theta-h_{33}\sin\theta\sin\phi-\sin\theta\sin\phi)
\end{eqnarray}
Then, based on Eq. (\ref{126}) and Eq. (\ref{127}), we get
\begin{equation}
\dot{x}=\frac{d_2^{'} a_1^{'}-d_1^{'}a_2^{'}}{a_2^{'} b_1^{'}-a_1^{'} b_2^{'}}\dot{z}+\frac{c_1 a_2^{'}-c_2a_1^{'}}{a_2^{'} b_1^{'}-a_1^{'} b_2^{'}}\label{128}
\end{equation}
\begin{equation}
\dot{y}=\frac{d_2^{'} b_1^{'}-d_1^{'}b_2^{'}}{a_1^{'} b_2^{'}-a_2^{'} b_1^{'}}\dot{z}+\frac{c_1 b_2^{'}-c_2b_1^{'}}{a_1^{'} b_2^{'}-a_2^{'} b_1^{'}}\label{129}
\end{equation}
With Eq. (\ref{122}) and Eq. (\ref{125}), we get
\begin{equation}
a^2(-\cos\theta\dot{\eta}+h_{33}\dot{z}+h_{13}\dot{x}+h_{23}\dot{y}+\dot{z})=c_3\label{130}
\end{equation}
then
\begin{equation}
a^2\dot{\eta}(-\cos\theta+h_{33}\frac{dz}{d\eta}+h_{13}\frac{dx}{d\eta}+h_{23}\frac{dy}{d\eta}+\frac{dz}{d\eta})=c_3\label{131}
\end{equation}
Due to the fact that $|h_{33}\frac{dz}{d\eta}+h_{13}\frac{dx}{d\eta}+h_{23}\frac{dy}{d\eta}|\ll|-\cos\theta+\frac{dz}{d\eta}|$ when $\theta$ is not near $0$, thus,
\begin{equation}
a^2\dot{\eta}=a^2\frac{d\eta}{dp}=c_4\label{132}
\end{equation}
With
\begin{equation}
\frac{dz}{d\eta}\approx1\label{133}
\end{equation}
we get
\begin{equation}
a^2\frac{dz}{dp}=a^2\frac{dz}{d\eta}\frac{d\eta}{dp}=c_4\label{134}
\end{equation}

Then, Eq. (\ref{128}) and Eq. (\ref{129}) change to
\begin{eqnarray}
\frac{dx}{dz}=\frac{d_2 a_1-d_1a_2}{a_2 b_1-a_1 b_2}+\frac{c_1 a_2-c_2a_1}{a_2 b_1-a_1 b_2}\\
\frac{dy}{dz}=\frac{d_2 b_1-d_1b_2}{a_1 b_2-a_2 b_1}+\frac{c_1 b_2-c_2b_1}{a_1 b_2-a_2 b_1}
\end{eqnarray}
where $c_1$ and $c_2$ are two new constants and
\begin{eqnarray}
a_1&=&h_{22}\cos\phi+\cos\phi-h_{12}\sin\phi\\
a_2&=&h_{22}\cos\theta+\cos\theta-h_{23}\sin\theta\sin\phi\\
b_1&=&h_{12}\cos\phi-h_{11}\sin\phi-\sin\phi\\
b_2&=&h_{12}\cos\theta-h_{13}\sin\theta\sin\phi\\
d_1&=&h_{23}\cos\phi-h_{13}\sin\phi\\
d_2&=&h_{23}\cos\theta-h_{33}\sin\theta\sin\phi-\sin\theta\sin\phi
\end{eqnarray}

\section*{Appendix c}
Based on $\frac{dx}{dz}$ and $\frac{dy}{dz}$ derived in APPENDIX B, we can express $\int_{-L}^{-AL}\frac{dx}{dz}$ and $\int_{-L}^{-AL}\frac{dy}{dz}dz$ as $\int_{-L}^{-AL}\frac{dx}{dz}dz=e_1+c_1 f_1+c_2 g_1$ and $\int_{-L}^{-AL}\frac{dy}{dz}dz=e_2+c_1 f_2+c_2 g_2$, respectively. The parameters $e_1$, $f_1$, $g_1$, $e_2$, $f_2$ and $g_2$ are given as follows
\begin{eqnarray}
e_1=\int_{-L}^{-AL} \frac{d_2 a_1-d_1a_2}{a_2 b_1-a_1 b_2}dz\\
f_1=\int_{-L}^{-AL} \frac{a_2}{a_2 b_1-a_1 b_2}dz\\
g_1=\int_{-L}^{-AL} \frac{-a_1}{a_2 b_1-a_1 b_2}dz\\
e_2=\int_{-L}^{-AL} \frac{d_2 b_1-d_1b_2}{a_1 b_2-a_2 b_1}dz\\
f_2=\int_{-L}^{-AL} \frac{b_2}{a_1 b_2-a_2 b_1}dz\\
g_2=\int_{-L}^{-AL} \frac{-b_1}{a_1 b_2-a_2 b_1}dz
\end{eqnarray}

Similarly, $\int_{-AL}^0\frac{dx}{dz}$ and $\int_{-AL}^0\frac{dy}{dz}dz$ can be expressed as $\int_{-AL}^0\frac{dx}{dz}dz=h_1+c_1 i_1+c_2 j_1-\alpha_x AL$ and $\int_{-AL}^0\frac{dy}{dz}dz=h_2+c_1 i_2+c_2 j_2-\alpha_y AL$, respectively. The parameters $h_1$, $i_1$, $j_1$, $h_2$, $i_2$ and $j_2$ are given as follows
\begin{eqnarray}
h_1&=&\int_{-AL}^0 \frac{d_2 a_1-d_1a_2}{a_2 b_1-a_1 b_2}dz\\
i_1&=&\int_{-AL}^0 \frac{a_2}{a_2 b_1-a_1 b_2}dz\\
j_1&=&\int_{-AL}^0 \frac{-a_1}{a_2 b_1-a_1 b_2}dz\\
h_2&=&\int_{-AL}^0 \frac{d_2 b_1-d_1b_2}{a_1 b_2-a_2 b_1}dz\\
i_2&=&\int_{-AL}^0 \frac{b_2}{a_1 b_2-a_2 b_1}dz\\
j_2&=&\int_{-AL}^0 \frac{-b_1}{a_1 b_2-a_2 b_1}dz
\end{eqnarray}

\bsp	% typesetting comment
\label{lastpage}
\end{document}